\documentclass{epl}

\title{Fast projectile stopping power of\\ two-dimensional strongly correlated electron liquids}
\shorttitle{Stopping power in 2D electron liquids}

\author{D. Ballester \and A. M. Fuentes \and I. M. Tkachenko}
\shortauthor{D. Ballester \etal}

\institute{                    
Departament de Matem\`atica Aplicada, Universitat Polit\`ecnica de Val\`encia - 46022 Valencia, Spain
}

\begin{document}

\maketitle

\begin{abstract}
We study the high-velocity-projectile limit of the polarizational contribution to the in-plane stopping power in a strongly coupled two-dimensional electron liquid. The dielectric formalism based on the method of moments is employed. The frequency moments of the loss function are calculated using the model Hamiltonian including the two-dimensional Coulomb interaction potential proportional to the inverse power of $k$. We prove that the leading term of the high-velocity asymptote, like in the {\it random-phase approximation}, is not affected by correlations.
\end{abstract}

In the last two decades, there has been a great interest in the investigation of static and dynamic properties of two-dimensional charged particle systems. Two-dimensional electron liquids trapped above the free surface of liquid helium \cite{GA76} constitutes an example of this kind of systems within the classical regime.

Moreover, recent advances in modern semiconductor nanotechnology have made it possible
to fabricate quasi-two-dimensional quantum well structures of electron
layers conforming a degenerate strongly correlated liquid phase. This is the case, for instance, of the systems consisting of electrons confined in the vicinity of a junction between a semiconductor and insulators \cite{Abrahams}, e.g., in a metal-oxide-semiconductor field-effect transistor, MOSFET, structure. Here, quasi-two-dimensional means that electrons have quantized energy levels along one dimension, whereas they are free to move within the
two-dimensional junction.

The system in question is a two-dimensional negatively charged one component plasma (OCP) immersed
in a uniform and rigid neutralizing background of positive charges. We presume that electrons interact via a two-dimensional Coulomb-like potential, inversely
proportional to the distance.

There are two dimensionless parameters, which usually characterize these systems: the first one is used to quantify its degeneracy: $D =\beta E_{\ab F}$, where $\beta $ is the inverse temperature in energy units and $E_{\ab F}=\hbar ^{2} k_{\ab F}^{2} /2m$ is the Fermi energy, $k_{\ab F}=\sqrt{2 \pi n}$ being the Fermi wave-vector, $m$ the electron mass and $n$ stands for the two-dimensional electron density; the second dimensionless parameter is the coupling parameter, $\Gamma =\beta  e^{2}/a$, where $a=\left( \pi n\right) ^{-1/2}$ is the 2D Wigner-Seitz radius and $e$ the electron charge. Whereas the parameter $\Gamma$ characterizes classical Coulomb systems, for quantal systems we might replace the temperature, $\beta^{-1}$, by $E_{\ab F}$. In this case $\Gamma$ equals the Brueckner parameter, $r_{\ab s}=a/a_{\ab B}$, $a_{\ab B}$ being the Bohr radius.

Two-dimensional electron liquids are known to crystallize at about $\Gamma \approx 137\pm 15$ in the classical case \cite{GA79}, and at $r_{\ab s}\approx 37\pm 5$ for the zero-temperature system \cite{TC89}, although this value can be affected by the presence of disorder \cite{CT95}. For Coulomb systems, the strongly coupled regime is characterized by a value of the coupling parameter much higher than unity. Nowadays, this kind of experimental physical conditions have become usual. In this situation, interparticle correlational effects become crucial and, consequently, to describe such systems theoretically, we need to go beyond the \textit{random-phase approximation} (RPA). 

In general, interaction of charged particles with condensed matter can be studied by means of the system stopping power. This function might be helpful in order to derive general diagnostic tools for this kind of physical systems. The stopping power accounts for the energy lost by an external positively charged projectile (which is assumed to move along a classical trajectory) as it passes through matter. There are several ways of dealing with this problem. Within the linear dielectric approach \cite{Lindhard,BD93} one assumes the incident projectile to have a constant velocity, and its energy loss is related to the strength of the dipole created between the ion and the center of charge of the induced charge density. This physical picture becomes more accurate for high-velocity and heavy projectiles. On the other hand, under the scattering picture \cite{N95,KP95,BNE97} one relates the stopping power to the transfer of momentum produced by electron collisions with an effective potential created by the moving ion. Further works have dealt also with nonlinear screening effects by means of the quadratic response approach within the RPA \cite{BPE99} and the employment of the density functional theory \cite{ZNE2005}.

In the sequel we will focuse on the part of the energy losses due to linear polarization of the
medium. It was Lindhard \cite{Lindhard} who derived a general expression
relating the imaginary part of the inverse dielectric function to the
polarizational losses valid in three-dimensional systems. Thus, in this
context, the derivation of a consistent dielectric formalism has played a key
role.

More recently, Bret and Deutsch \cite{BD93} obtained a similar
expression for the in-plane polarizational stopping power in a
two-dimensional Coulomb system. For a point-like charged particle this
expression reads as

\begin{equation}
-\frac{\upd E}{\upd x}=\frac{2\left( Ze\right) ^{2}}{\pi v}\int\limits_{0}^{\infty
}\upd k\int\limits_{0}^{kv}\upd \omega \frac{\omega }{\sqrt{(kv)^{2}-\omega ^{2}}}%
{\rm Im}\left( -\frac{1}{\varepsilon \left( k,\omega \right) }\right) .
\label{BD1}
\end{equation}

Moreover, these authors derived a dielectric function within the RPA and
obtained asymptotic expressions for the polarizational stopping power for high- and low-velocity projectiles. In the former case, in contrast to the well-known 3D asymptote \cite{Bethe,Bloch,Larkin}, it was shown in Ref. \cite{BD93} that the leading term of the stopping power is inversely proportional to the projectile velocity. This result has been later confirmed by means of different approaches \cite{N95,KP95,BNE97,WM95,WM97}. The influence of electronic correlations in a zero-temperature electron gas on the low-velocity asymptote was also considered by Wang and Ma \cite{WM95} who used a parametrized expression for the static local-field correction factor. However, these authors explicitly neglected the effect of electronic correlations on the high-velocity limiting stopping power, an assumption which was entirely based on a physical picture. Our goal here is to provide rigorous argument supporting this physically sound assumption.

More precisely, our dielectric formalism, which is based on the method of moments, permits us to account for correlational contributions and handle the problem analitically in order to show that the first term of the asymptotic expansion of the two-dimensional stopping power is unaffected by electron correlational effects.

\section{Dielectric formalism}

The (truncated) method of moments provides a suitable mathematical tool in
order to reconstruct response functions or susceptibilities of a
non-perturbative physical system from its known frequency (power) moments or
sum rules \cite{AMT85}. In particular, for the inverse dielectric function we refer to
the frequency moments of its imaginary part (we assume the system to be
isotropic),%
\begin{equation}
C_{\nu }=-\frac{1}{\pi }\int_{-\infty }^{\infty }\omega ^{\nu -1}%
{\rm Im}\varepsilon ^{-1}\left( k,\omega \right) \upd \omega ,\quad \nu
=0,1,\ldots .  \label{Cnu}
\end{equation}

Here, odd moments ($\nu =2n+1$, $n=0,1,\ldots$) vanish due to the parity of the loss function, $-{\rm Im}\varepsilon ^{-1}\left( k,\omega \right)/\omega$. In addition, only the moments $C_{0}$, $C_{2}$, and $C_{4}$ converge, in general.
These are related to the static properties of the electron system \cite{AMT85,OT92}:
\begin{equation}
C_{0}\left( k\right) =1-\varepsilon ^{-1}\left( k,0\right) ,  \label{C0}
\end{equation}%
\begin{equation}
C_{2}\left( k\right) =\frac{2\pi ne^{2}}{m}k\equiv \omega _{\ab{2D}}^{2},
\label{C2}
\end{equation}%
$\omega _{\ab{2D}}=\omega _{\ab{2D}}\left( k\right) $ being the 2D plasma frequency,%
\begin{equation}
C_{4}\left( k\right) =\omega _{\ab{2D}}^{4}\left( 1+K\left( k\right) +L\left(
k\right) \right) ,  \label{C4}
\end{equation}%
where%
\begin{equation}
K\left( k\right) =\frac{3E_{\ab{kin}}}{2\pi ne^{2}}k+\frac{\hbar ^{2}}{8\pi
ne^{2}m}k^{3}  \label{K}
\end{equation}%
is the kinetic contribution to the fourth sum rule, $E_{\ab{kin}}$ being the
average kinetic energy per electron, which, for example, will coincide with $\beta ^{-1}$ for a classical system. Furthermore, the correlation contribution reads as%
\begin{equation}
L\left( k\right) =\frac{1}{N}\sum_{\mathbf{q}(\neq 0,\neq \mathbf{k})}\frac{%
\left( \mathbf{k}\cdot \mathbf{q}\right) ^{2}}{k^{3}q}\left[ S\left( \mathbf{%
k}+\mathbf{q}\right) -S\left( q\right) \right] ,  \label{L}
\end{equation}%
which is applicable for the two-dimensional Coulomb system. It must be stressed that, whereas the sum rules $C_{2}$ and $C_{4}$ are obtained by the usual derivation procedure from the Kubo formula, the moment $C_{0}$ stems from the Riesz-Herglotz transformation \cite{Akh,KN} for a non-magnetized plasma \cite{AMT85}.

Further, owing to the Kramers-Kronig relations and the Nevanlinna formula \cite{Akh, KN}, the expression%
\begin{equation}
\varepsilon ^{-1}\left( k,z\right) =1+\frac{\omega _{\ab{2D}}^{2}\left( k\right) %
\left[ z+q\left( k,z\right) \right] }{z\left[ z^{2}-\omega _{2}^{2}\left(
k\right) \right] +q\left( k,z\right) \left[ z^{2}-\omega _{1}^{2}\left(
k\right) \right] }  \label{nevanlinna}
\end{equation}%
establishes a one-to-one correspondence between the set of all possible dielectric
(complex) functions satistying the sum rules (\ref{C0}-\ref{C4}), and an arbitrary
parameter function $q\left( k,z\right) $. The latter must be analytic in the upper
half-plane ${\rm Im}z>0$ and possess a positive imaginary part there, and, in addition,
it must fulfill the condition: $\lim_{z\rightarrow \infty }q\left( k,z\right) /z=0$\ for
$\theta <\arg (z)<\pi -\theta $ ($0<\theta <\pi /2$).

The system parameters appearing in (\ref{nevanlinna}) depend only on the frequency
moments:%
\begin{equation}
\omega _{1}^{2}\left( k\right) =C_{2}/C_{0}=\omega _{\ab{2D}}^{2}\left( k\right)
\left( 1-\varepsilon ^{-1}\left( k,0\right) \right) ^{-1},  \label{omega1}
\end{equation}%
\begin{equation}
\omega _{2}^{2}\left( k\right) =C_{4}/C_{2}=\omega _{\ab{2D}}^{2}\left( k\right)
\left( 1+K\left( k\right) +L\left( k\right) \right) .  \label{omega2}
\end{equation}

The function $q\left( k,z\right) $ is phenomenologically related to the
damping of the plasmon mode of the system. Actually, if we put $q\left(k,z\right) =i0^{+}$, we obtain for the loss function:%
\begin{equation}
-\frac{{\rm Im}\varepsilon ^{-1}\left( k,\omega \right) }{\omega }=\pi 
\left[ \frac{\omega _{\ab{2D}}^{2}}{2\omega _{2}^{2}}\delta \left( \omega +\omega
_{2}\right) +  \frac{\left( \omega _{2}^{2}-\omega _{1}^{2}\right) }{
\omega _{2}^{2}} C_{0} \delta \left( \omega \right)  +   \frac{\omega _{\ab{2D}}^{2}}{%
2\omega _{2}^{2}}\delta \left( \omega -\omega _{2}\right) \right] ,
\label{canonical}
\end{equation}%
which resembles the Feynman-like approximation for the dynamic structure
factor. This expression obviously belongs to the set of solutions of the
(truncated) moment problem, i.e., it does satisfy all three frequency sum
rules (\ref{C0}-\ref{C4}). In a mathematical context, the function (\ref{canonical}) is usually referred to as a canonical solution of the moment problem.

It must be pointed out that, unlike (\ref{canonical}), the model expressions of Eq. (34) in \cite{BD93} and Eq. (23) in \cite{WM95} fail to
satisfy the sum rules (\ref{C0}) and (\ref{C4}). In addition, for a strongly coupled system, $\Gamma \gg 1$, one might
neglect the kinetic contribution, $K\left( k\right) $, in the moment $C_{4}\left( k\right) $. This leads to the inverse dielectric function obtained by Golden and Kalman by means of the QLC approximation \cite{QLCA}.

As the Feynman-like approximation,  expression (\ref{canonical}) reproduces a collective excitation at the frequency $\omega_{2}\left( k\right)$. The additional central peak here permits to account for hydrodynamic difussional effects \cite{DK2003}. Whereas in a multicomponent plasma this peak becomes dominant at low frequencies, in case of a one component plasma it vanishes as $k$ decreases \cite{DK2001,HMDP75}, as it can be observed from (\ref{canonical}) just by taking the long wave-length limit and recalling the well-known compressibility sum rule (see also expression (\ref{omega2_ksmall}) below). In fact, this feature is related to the existence of a vanishing static (internal) conductivity in the long wave-length limit and might be understood as a macroscopic manifestation of the conservation of charged momentum during interparticle collisions at a microscopic level, which is obviously not applicable in a multicomponent system.

\section{Fast projectile polarizational stopping power}

Here we are interested in the fast projectile stopping power of Coulomb systems with well-defined collective modes. Then it is sufficient to consider the canonical solution or the loss function in the form of (\ref{canonical}). For a weakly coupled OCP, the coupling parameter, $\Gamma$ or $r_{\ab s}$, is $\ll 1$. Under such conditions one might make use of
the RPA dispersion law which implies neglecting particle correlational contributions to the collective excitation frequency. Then,%
\begin{equation}
\omega _{2}^{2}=\omega _{\ab{2D}}^{2}\left[ 1+\frac{3E_{\ab{kin}}^{\ab{RPA}}}{2\pi ne^{2}}k+\frac{%
\hbar ^{2}}{8\pi ne^{2}m}k^{3}\right] .  \label{resonance1}
\end{equation}

Further, as a first approximation, one might consider only the asymptotic
leading terms in the previous expression for the limits of large and small
wave-vectors. Thus,%
\begin{equation}
\omega _{2}^{2}\approx \omega _{\ab{2D}}^{2}\left[ 1+\frac{\hbar ^{2}}{8\pi
ne^{2}m}k^{3}\right] .  \label{resonance2}
\end{equation}

By substituting expressions (\ref{canonical}) and (\ref{resonance2}) into (\ref{BD1}), one obtains%
\begin{equation}
-\frac{\upd E}{\upd x}=\frac{2\sqrt{2}Z^{2}e^{2}}{r_{\ab s}a_{\ab B}^{2}}\int_{z_{\min
}}^{z_{\max }}\frac{b^{2}}{\sqrt{1-\delta b^{2}z^{-1}-b^{2}z^{2}}}\upd z,
\label{BD2}
\end{equation}%
where we introduce the parameter $\delta =(2k_{\ab F}a_{\ab B})^{-1}$ and the
new variable $b= v_{\ab F}/v$. The polynomial equation \cite{WM95} $z^{3}-b^{-2}z+\delta =0$ has one negative and two positive roots, $z_{\min}$, $z_{\max }$, $z_{\min}<z_{\max }$, if

\begin{equation}
b\ll \delta^{-2/3}. \label{conditionb}
\end{equation}

The leading term of the asymptotic expansion of (\ref{BD2}) can then be calculated exactly:

\begin{equation}
-\frac{\upd E}{\upd x}=\frac{\pi \sqrt{2}Z^{2}e^{2}}{r_{\ab s}a_{\ab B}^{2}}  \frac{%
v_{\ab F}}{v} + \mathcal{O}\left( \frac{v_{\ab F}^{3}}{v^{3}}\right)  .
\label{asymptotic1}
\end{equation}%
This result was initially obtained by Bret and Deutsch \cite{BD93}.

On the other hand, if we want to consider all contributions in (\ref{resonance1}) and the effects of an arbitrary coupling in the electron gas, i.e., to go beyond the RPA, we should consider $\omega_{2}^{2}\left( k\right) $  as given by (\ref{omega2}). In our case, it suffices to consider the expansion of the expression (\ref{omega2}) into the power series in the wave-vector. For small $k$, we get \cite{QLCA}%
\begin{equation}
\omega _{2}^{2}\left( k\downarrow 0\right) \simeq \omega _{\ab{2D}}^{2}\left[ 1+%
\frac{3E_{\ab{kin}}}{2\pi ne^{2}}k+\frac{5E_{\ab{c}}}{16\pi ne^{2}}k+\mathcal{O}\left(
k^{2}\right) \right] ,  \label{omega2_ksmall}
\end{equation}%
$E_{\ab{c}}$ being the correlation energy per particle. Similarly, for large $k$ \cite{QLCA}:%
\begin{equation}
\omega _{2}^{2}\left( k\uparrow \infty \right) \simeq \omega _{\ab{2D}}^{2}\left[
\frac{\hbar ^{2}}{8\pi ne^{2}m}k^{3}+\frac{3E_{\ab{kin}}}{2\pi ne^{2}}k+\mathcal{O%
}\left( 1\right) \right] .  \label{omega2_klarge}
\end{equation}

Thus, we might make use of the interpolation formula:%
\begin{equation}
\omega _{2}^{2}\left( k\right) \approx \omega _{\ab{2D}}^{2}\left[ 1+\frac{%
3E_{\ab{kin}}}{\pi ne^{2}}k+\frac{5E_{\ab{c}}}{16\pi ne^{2}}k+\frac{\hbar ^{2}}{8\pi
ne^{2}m}k^{3}\right] .  \label{omega2_interpol}
\end{equation}

Again, by inserting expression (\ref{canonical}), now with (\ref{omega2_interpol}) into formula (\ref{BD1}) we get%
\begin{equation}
-\frac{\upd E}{\upd x}=\frac{2\sqrt{2}Z^{2}e^{2}}{r_{\ab s}a_{\ab B}^{2}}\int_{z_{\min
}}^{z_{\max }}\frac{b^{2}}{\sqrt{1-\xi b^{2}-\delta b^{2}z^{-1}-b^{2}z^{2}}}%
\upd z,  \label{BD3}
\end{equation}%
where the term
\begin{equation}
\xi =3\frac{E_{\ab{kin}}}{E_{\ab F}}+\frac{5}{16}\frac{E_{\ab{c}}}{E_{\ab F}}  \label{xi}
\end{equation}%
accounts for the contribution of the average kinetic and correlation
energies per electron. Clearly, this integral can be recast into%
\begin{equation}
-\frac{\upd E}{\upd x}=\frac{2\sqrt{2}Z^{2}e^{2}}{r_{\ab s}a_{\ab B}^{2}}\sqrt{1-\xi b^{2}}%
\int_{\bar{z}_{\min }}^{\bar{z}_{\max }}\frac{\bar{b}^{2}}{\sqrt{1-\delta
\bar{b}^{2}z^{-1}-\bar{b}^{2}z^{2}}}\upd z,  \label{BD4}
\end{equation}%
with

\begin{equation}
\bar{b}^{2}=\frac{b^{2}}{1-\xi b^{2}}.  \label{bbar}
\end{equation}

Notice that in this case the condition (\ref{conditionb}) transforms into:

\begin{equation}
b\ll \frac{1}{\sqrt{\xi+\delta^{4/3}}}.  \label{conditionbbar}
\end{equation}

The leading term of the asymptotic expansion of (\ref{BD4}) for $b\rightarrow 0$ obviously coincides with that of (\ref{asymptotic1}). This remains true even if we abandon the singular form of the canonical solution (\ref{canonical}) and substitute the $\delta$-functions by adequate Gaussian densities, thus taking into account an exponentially small damping of the plasmon mode. Indeed, the form (\ref{BD4}) is recovered then by a straightforward application of the Laplace method of asymptotic analysis.

Thus we prove that for the case of high-velocity projectiles, the correction to the stopping power due to correlational effects does not affect the main asymptotic term as $v/v_{\ab F}\rightarrow \infty$, obtained in Ref. \cite{BD93}. Finite damping \cite{QLCA} can be taken into account using a non-canonical solution of the moment problem (\ref{nevanlinna}).

Similar considerations are valid also for the 3D interacting electronic gas, where the Bethe-Bloch-Larkin \cite{Bethe,Bloch,Larkin} asymptote is not affected by correlations \cite{OT2001}.

\acknowledgments
I.M.T. gratefully acknowledges the financial support provided in the framework of the GSI-INTAS project No. 03-54-4254.

\end{document}